\documentclass[prl,aps,twocolumn,superscriptaddress,showpacs,amsmath,amssymb,floatfix]{revtex4}

\usepackage{graphicx}

\newcommand{\be}{\begin{equation}}
\newcommand{\ee}{\end{equation}}

\begin{document}
\title{Superfluid Fermi gas in a 1D optical lattice}

\author{G. Orso}
\affiliation{BEC-INFM and Dipartimento di Fisica, Universita' di Trento, 
1-38050 Povo, Italy}
\author{G.V. Shlyapnikov}
\affiliation{\mbox{Laboratoire Physique Theorique et Modeles Statistiques,
Universite' Paris Sud, Bat. 100, 91405 Orsay Cedex, France}}
\affiliation{\mbox{Van der Waals-Zeeman Institute, University of Amsterdam,
Valckenierstraat 65/67, 1018 XE Amsterdam, The Netherlands}}

\begin{abstract}
We calculate the superfluid transition
temperature for a two-component 3D Fermi
gas in a 1D tight optical lattice and discuss a dimensional
crossover from the 3D to quasi-2D regime. For the geometry of finite size discs
in the 1D lattice, we find that even for a large number of atoms
per disc, the critical effective tunneling rate 
for a quantum transition to the Mott insulator state can be
large compared to the loss rate caused by three-body
recombination. This allows the observation of the
Mott transition, in contrast
to the case of Bose-condensed gases in the same geometry.
\end{abstract}
\maketitle

The observation of a BCS superfluid transition remains a
challenging goal in the studies of ultracold Fermi gases.
It was recently suggested that gases confined to low dimensions
are promising candidates for achieving superfluidity as the
confinement enhances interaction effects \cite{gora}. Adding a tunable
periodic potential allows one to combine the benefit of the
reduced dimensionality with the advantage to work with 
large yet coherent samples. In particular, it has recently been shown that
the study of the center of mass oscillations of the cloud
in a 1D lattice plus a superimposed weak harmonic potential allows one to 
probe the superfluid transition \cite{umklapp}.  
Importantly, the presence of periodic
potential introduces a much richer physics related to a possibility
of observing a variety of quantum phase transitions \cite{greiner}.

In this Letter we obtain the BCS transition temperature $T_c$
for a two-component Fermi gas in a 1D optical lattice, assuming that
the Fermi energy is small compared to the interband gap and, hence,
superfluid pairing occurs only in the lowest band. 
This requires us to reveal how the presence of the 1D lattice 
renormalizes the effective coupling constant at the Fermi level.
For the geometry
of finite size discs in the 1D lattice, we also discuss the possibility
of achieving the Superfluid-Mott Insulator quantum transition by tuning the
lattice depth above a critical value. In this peculiar phase, the gas is 
superfluid in each separate disc
but the coherence along the lattice direction is completely lost.  We show
that for Fermi superfluids the critical effective tunneling rate can be large 
compared to the loss rate of all inelastic processes and therefore
the Mott transition can be achieved. This result is a direct 
consequence of the Fermi statistics and is in marked contrast with the
case of Bose-Einstein condensates in the same geometry, where 
the Mott transition can be hardly observed as pointed out by
Hadzibabic et al. \cite{dalibard},  
unless the number of atoms per disc is very small.

We consider a two component atomic Fermi gas in the presence of a
one dimensional (1D) optical potential 
\be
V_{opt}=sE_R\sin^2 q_Bz,
\label{V}
\ee
where $s$ is a dimensionless parameter coming from the
intensity of the laser beam, $E_R  = \hbar^2q_B^2/2m$ is the recoil 
energy, with $\hbar q_B$ being the Bragg momentum and $m$ the atom mass. 
The potential (\ref{V}) 
has periodicity $d=\pi /q_B$  along the $z$-axis.
The weak attraction between atoms in different internal states 
is modeled  by a $s$-wave pseudopotential 
$U(\mathbf r)=g \delta(\mathbf r)\partial_r(r \cdot)$ with
 coupling constant $g=4\pi \hbar^2 a/m$, where $a<0$ is the 3D 
scattering length.

We will discuss the situation where the laser intensity is sufficiently 
large $(s \gtrsim 5)$ and the Fermi energy $\epsilon_F$ is small 
compared to the interband gap $\epsilon_g$. 
 We thus confine ourselves to the lowest Bloch band where the physics is governed by the ratio of the Fermi energy
to the bandwidth $4t$, 
 where $t$ is the hopping rate between neighboring wells. For 
$\epsilon_F <4t$ the Fermi surface is closed and the system 
retains a 3D behaviour, whereas in the case of $\epsilon_F >4t$ 
the Fermi surface is open and the system undergoes a dimensional crossover.
Hence, one has two distinct regimes: an anisotropic 3D regime
($\epsilon_F\ll t$) and a quasi-2D regime ($\epsilon_g \gg 
\epsilon_F\gg t$). This is clearly different from the case of a 3D 
lattice \cite{zoll}
where the Fermi energy scales with the bandwidth
and can therefore be much smaller than the corresponding  value 
in free space for a given atom density.


The mean field transition temperature $T_c^0$ is the highest 
temperature at which the Gorkov equation for the gap 
parameter has a non-trivial solution \cite{degennes}. 
This gives
\be\label{first}
\frac{1}{g_{eff}}=\int \frac{d^3 q}{(2\pi)^3} {\textrm P}\frac{1}{\xi_{\mathbf q}}\frac{1}{\exp(\xi_\mathbf q/T_c^0)+1},
\ee
where $g_{eff}$ is an effective coupling constant.
The symbol P stands for Principal value and
$\xi_\mathbf q=\hbar^2 \mathbf q_\perp^2/2m +\epsilon_1(q_z)-\mu$,
where $\mathbf q_{\perp}$ is the momentum in the direction perpendicular to
the lattice, $\epsilon(q_z)$ is the
band dispersion and $\mu\simeq\epsilon_F$ is the chemical potential.
A straightforward integration of  Eq.(\ref{first}) yields
\be\label{tc0}
T_c^0=\frac{2 \gamma}{\pi} \mu 
\exp\left(\frac{1}{g_{eff}} \frac{1}{\nu(\mu)}-F(\mu)\right),
\ee
with $\gamma=1.781$ and
$\nu(\mu)=\int \delta(\xi_\mathbf q)d\mathbf q /(2\pi)^3$ being the density of 
states per internal state at the Fermi level. The function $F$ is defined as 
\be\label{f}
F=-\frac{\int_{-q_B}^{q_B}dq_z \ln\left(1-\epsilon (q_z)/\mu)\Theta(\mu-\epsilon(q_z)\right)} 
{\int_{-q_B}^{q_B}dq_z \Theta(\mu-\epsilon(q_z))},
\ee
where $\Theta(x)$ is the unit-step function. 

The effective coupling constant is related to the scattering 
amplitude $f(E)$ for Cooper pairs by 
$g_{eff}^{-1}=(m/4\pi \hbar^2)\textrm{Re}[1/f(E=2 \mu)]$ \cite{abrikosov}. 
This requires us to solve the two-body problem for finding
the scattering amplitude in the presence of the 1D lattice.
In this case the expression for $f(E)$ is given by
\be\label{scat1} 
f(E)=a \int dZ \phi^*_E(Z,0) \partial_r (r \Psi(Z,\mathbf r))_{r=0}
\ee
where $\phi_E(Z, \mathbf r)=\phi_{1 q_{z}}(z_1)\phi_{1 -q_{z}}(z_2)e^{i \mathbf q_\perp \mathbf r_\perp}$
is the incoming wavefunction for two atoms undergoing Cooper pairing.
The center of mass and relative coordinates
are $Z=(z_1+z_2)/2$ and $\mathbf r=\mathbf r_1-\mathbf r_2$, 
and $E=\hbar^2 \mathbf q_\perp^2/m+2\epsilon_1(q_z)$ is the total energy. 
The two-particle wavefunction  $\Psi(Z,\mathbf r)$ 
obeys the Schr\"odinger equation
\be
\!\!\left(\!\!-\frac{\hbar^2}{m}\Delta-\frac{\hbar^2}{4m}\frac{\partial^2}{\partial Z^2}+V(Z,z) 
+g\delta(\mathbf{r})\frac{\partial }{\partial r} r\!-\!E\!\right)\!
\Psi\!=0,\!
\label{schrodinger}
\ee
where $V(Z,z)=V_{opt}(z_1)+V_{opt}(z_2)$. 
The solution of Eq.(\ref{schrodinger}) can be written as
\begin{eqnarray}    
\Psi(Z,\mathbf r)=\phi_E(Z, \mathbf r)+g\int dZ 
G_E \partial_{r^\prime}
(r^\prime \Psi(\mathbf 
r^\prime,Z^\prime))_{r^\prime=0}  \label{Psi} 
\end{eqnarray}
where $G_E(\mathbf{r},Z;\mathbf{0},Z^{\prime})$ is the Green function
of Eq.(\ref{schrodinger}) with $g=0$.
The behaviour of the Green function at short distances $r$  
is governed by the Laplacian term in Eq.(\ref{schrodinger})
yielding $G_E(\mathbf{r},Z;\mathbf{0},Z^{\prime})=
-\delta(Z-Z^\prime)m/4\pi \hbar^2 r+K_E(Z,Z^\prime)$,
where $K_E(Z,Z^\prime)$ is a regular function. 
Then, from Eq.(\ref{Psi}) we immediately obtain an equation 
for the function $Y(Z)=\partial_{r}(r \Psi(\mathbf 
r,Z))_{r=0}$ appearing in Eq.(\ref{scat1}): 
\be\label{nin}
Y(Z)=\phi_E(Z,0)+g \int dZ^\prime K_{E}(Z,Z^\prime)Y(Z^\prime). 
\ee
Writing the kernel of the integral equation (\ref{nin}) in the form
$K_E(Z,Z^{\prime })=[G_E(\mathbf{r},Z;\mathbf{0},Z^{\prime})-
G_{E=0}(\mathbf{r},Z;\mathbf{0},Z^{\prime})]_{r=0}+K_{E=0}(Z,Z^\prime)$,
we expand the Green function $G_E$ in eigenstates
of non-interacting atoms: 
\begin{eqnarray}
 && G_E(\mathbf{r},Z;\mathbf{0},Z^{\prime}) =
  \sum_{n_1,n_2} \int\frac{d^{2}\mathbf{q}_{\perp }}{\left( 2\pi \right) ^{2}}
\int_{-q_B}^{q_B}\frac{dq_{1z}}{2\pi}\frac{dq_{2z}}{2\pi}
  e^{i\mathbf{q_\perp}\cdot \mathbf{r}} \notag\\
\times   && \frac{\phi_{n_1,q_{1z}}(Z)\phi_{n_2,q_{2z}}(Z)
\phi_{n_1,q_{1z}}^*(Z^\prime)\phi_{n_2,q_{2z}}^*(Z^\prime)}
  {E+i 0 -\epsilon_{n_1}(q_{1z})-\epsilon_{n_2}(q_{2z})-\hbar^2q_\perp^2/m},
  \label{GEexplicit}
\end{eqnarray}
and retain only the contribution of the lowest Bloch band. 
In the tight binding limit, these states 
can be written in terms of Wannier functions as 
$\phi_{1q_z}(z)\sim \sum_{\ell}e^{i\ell q_z d}w(z-\ell d)$, where 
$w(z)=(1/\pi^{1/4}\sigma^{1/2})\exp(-z^2/2\sigma^2)$
is a variational Gaussian ansatz. By minimizing the energy 
of non-interacting lattice atoms with respect to 
$\sigma$, one finds  $d/\sigma=\pi s^{1/4}\exp(-1/4\sqrt s)$ \cite{book}.

We now insert the ansatz $Y(Z)=A \sum_{\ell}w^2(Z-\ell d)$ 
into Eq.(\ref{nin}) and take into account that the relation 
$\int dZ^\prime K_{E=0}(Z,Z^\prime)Y(Z^\prime)dZ^\prime=
Y(Z) m/4\pi\hbar^2 a$ gives a critical  
value  of the scattering length $a=a_{cr}$ 
needed to form a two-body bound state in the lattice \cite{bstate}.
Then, using the dispersion relation 
$\epsilon_1(q_z)=2t(1-\cos(q_zd))$ and obtaining the kernel $K_E(Z,Z')$ 
on the basis of Eq.(\ref{GEexplicit}), 
we find the coefficient $A$ in
the expression for $Y(Z)$. Equation (\ref{scat1}) then leads
to the scattering amplitude 
\be\label{aaa}
f(E)=\frac{aC}{1-a/a_{cr}+(a /\sqrt{2 \pi}\sigma)\alpha(E/4t)}, 
\ee
where $C\!\!=\!\!d/\!\sqrt{2\pi}\sigma$. The function $\alpha(x)\!\!=\!\!i\!\arccos(1\!-\!x)$ for 
$x\!\!<\!\!2$, and $\alpha(x)\!=\!-\ln[x(1\!+\!\sqrt{1\!-\!2/x})^2/2]\!+\!i\pi$ for $x\!\geq\! 2$.  

Equation (\ref{aaa}) is one of the key results 
of this paper. It shows that the scattering amplitude 
undergoes a dimensional crossover as a function of energy. 
In the anisotropic 3D regime ($E \ll 8t$) 
we have $f=aC/(1-a/a_{cr}+iaC\sqrt{Em^*}/\hbar)$
where $m^*$ is the effective mass at the bottom of the band. 
In the quasi-2D regime ($E \gg 8t$), the tunneling between wells is 
irrelevant and
the two atoms are in the ground state of an effective harmonic potential 
of frequency $\omega_0=\hbar/m\sigma^2$. 
The scattering amplitude of Eq.(\ref{aaa}) should then reduce to
$f=f_{2D}d$ where 
$f_{2D}=(a/\sqrt{2\pi}\sigma)/\left(1+(a/\sqrt{2\pi}\sigma)
(\ln[\lambda \hbar \omega_{0}/E]+i \pi)\right)$ 
and $\lambda=0.915/\pi$ \cite{petrov}. 
This provides us with the asymptotic behaviour of
the critical value of the scattering length  
\be
\label{cr}
a_{cr}=-\sqrt{2\pi}\sigma \ln^{-1}
(\lambda \hbar \omega_0/2t),
\ee
which agrees with numerics of Ref.\cite{bstate}
already for $s \gtrsim 5$. 

We see that the 1D lattice affects $f(E)$ and the  coupling
constant $g_{eff}$ in a non-trivial way.
For $\epsilon_F < 4t$, the coupling is density independent 
and from Eq.(\ref{tc0}) we get 
\be\label{smaller}
T_c^0=\frac{2 \gamma}{\pi} e^{-F} \epsilon_F 
\exp \left[-\frac{\pi}{2 q_{zF} C}
\left(\frac{1}{|a|}-\frac{1}{|a_{cr}|}\right)\right],
\ee
where the function $F$ is given by Eq.(\ref{f}), and
$q_{zF}=\arccos(1-\epsilon_F/2t)/d$ is the Fermi wavevector
along the z-axis. Equation (\ref{smaller}) is valid provided $T_c^0 
\ll \epsilon_F$, which implies $|a|< |a_{cr}|$.
In the low density limit $\epsilon_F \ll 4t$, $T_c^0$ reduces to the
mean field transition temperature \cite{gorkov} for a homogeneous 
gas of atoms with an anisotropic quadratic dispersion 
and a renormalized 3D inverse scattering length 
$a_{eff}^{-1}=C^{-1}(|a|^{-1}-|a_{cr}|^{-1})$.
The presence of the lattice causes an 
effective shift 
of the resonance from $1/a=0$ to $1/a=1/a_{cr}<0$, which
in turn gives rise to a sharp increase in $T_c^0$ at a fixed 
value of the 3D scattering length.

For $\epsilon_F>4t$, the coupling constant $g_{eff}$ becomes density
dependent. Equation (\ref{f}) yields $F(\mu)=2 \ln[2/(1+\sqrt{1-4t/\mu})]$,
and from Eq.(\ref{tc0}) we find
\be\label{out}
T_c^0\!=\!
\frac{\gamma}{2 \pi}\!\left(\!1\!+\!\!\sqrt{1\!-\!\frac{4t}{\epsilon_F}}\right)
\sqrt{\epsilon_F 4t} 
\exp\!\left[\!\sqrt{\frac{\pi}{2}}\!\left(\!\frac{\sigma}{a}\!-\!\frac{\sigma}{a_{cr}}\!\right)\!\!\right].\!\!
\ee
Note that the exponent in the rhs
of Eq.(\ref{out}) does not depend on the Fermi energy, the
density of states being constant for $\epsilon_F>4t$.
For $\epsilon_F\gg 4t$, the ratio $\sqrt{4t/\epsilon_F}$ 
plays the role of a small parameter  
ensuring the inequality $T_c^0/\epsilon_F \ll 1$ also
for values of the scattering length 
larger than $a_{cr}$, but still $|a|\ll \sigma$.
In this regime, the system behaves as a
stack of quasi-2D superfluid gases weakly coupled by Josephson junctions. 
The transition temperature in each disc takes the form 
$T_c^0=\gamma\sqrt{2 \epsilon_F E_b}/\pi$ \cite{miyake}, where in our case $E_b$ is 
the binding energy of the two-body bound state in the 1D optical lattice. 
In the absence of coupling between the discs $(t=0)$, the gas in each disc is two-dimensional 
and the transition is therefore of the 
Kosterlitz-Thouless type. However, for 2D BCS superfluids a standard calculation of
the Kosterlitz-Thouless transition temperature gives a value that is lower than
$T_c$ by an amount $\sim T_c^2/\epsilon_F\ll T_c$ \cite{miyake} and thus lies inside a narrow region
of critical fuctuations in the neighborhood of $T_c$. Therefore the mean field approach leads
to a correct result for the transition temperature.   

We next proceed to evaluate Gorkov's correction to the transition temperature due to the polarization
of the medium \cite{gorkov}.  
Following Ref.\cite{pet}, we introduce the static Lindhard  function
\be\label{lind}
L(\mathbf p)=\int \frac{d{\mathbf q}}{(2\pi)^3} \frac{f(\xi_{\mathbf q})-
f(\xi_{\mathbf q +\mathbf p})}{\xi_{\mathbf q +\mathbf p}-\xi_{\mathbf q}},
\ee
where $f(x)=\Theta(-x)$ is the Fermi distribution 
at $T=0$. The induced interaction 
between two states $\mathbf q$ and $\mathbf q^\prime$ on the Fermi surface 
is given by $U_{ind}(\mathbf p)=g C L(\mathbf p)$, with 
$\mathbf p=\mathbf q+\mathbf q^\prime$ and $C$ defined as above.
Since $L(\mathbf p=0)=\nu(\mu)$, we write
$L(\mathbf p)=\nu(\mu) B(\mathbf p)$, where $B$ is a dimensionless
{\sl positive} function sensitive to the geometry of the Fermi surface.  
The critical temperature  is then given by 
$T_c=T_c^0 e^{-\langle B \rangle_{FS}}$, where 
\be\label{gorkov}
\langle B \rangle_{FS}=
\frac{\int B(\mathbf q+\mathbf q^\prime) \delta(\xi_{\mathbf q})
\delta(\xi_{\mathbf q^\prime})d\mathbf q d\mathbf q^\prime}
{\int \delta(\xi_{\mathbf q})\delta(\xi_{\mathbf q^\prime})d\mathbf q d\mathbf q^\prime}.
\ee
\begin{figure}[tb]
\begin{center}
\includegraphics[width=6cm,angle=270]{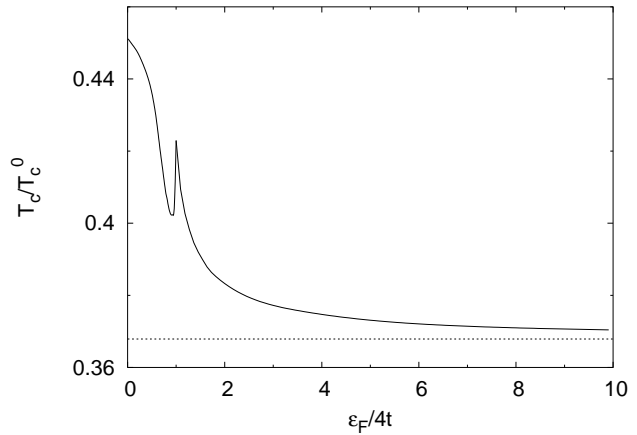}
\caption{Gorkov's correction versus $\epsilon_F/4t$. 
The limiting value $T_c/T_c^0=e^{-1}$ at $\epsilon_F/4t \gg 1$
is shown by a dotted line.}
\label{fig-gor}
\end{center}
\end{figure}
The integration in (\ref{gorkov}) is done numerically and 
the corresponding Gorkov 
correction $T_c/T_c^0$ is shown in Fig.\ref{fig-gor}. 
For $\epsilon_F \ll 4t$, the system has an anisotropic
 quadratic dispersion and we recover the result for the homogeneous case
$\langle B \rangle_{FS}=(1+2\ln 2)/3$, yielding 
$T_c=T_c^0/(4e)^{1/3}=0.45T_c^0$. In the 
limit $\epsilon_F \gg 4t$, the band dispersion 
$\epsilon(k_z)$ can be neglected in Eq.(\ref{lind})  and we find
 $T_c/T_c^0=e^{-1}$, in agreement with Ref.\cite{gora}. 
The cusp at $\epsilon_F=4t$ is expected 
as this is the point of the Van Hove singularity \cite{ash}
where the derivative of the density of states, 
$\partial \nu /\partial \epsilon$, diverges.

So far we have discussed the BCS superfluid transition
in a 1D optical lattice.   
In the second part of the Letter we assume that the superfluid gas 
is at zero temperature and it is confined in the $x,y$ directions by a trapping potential.
Then, as the tunneling rate between neighboring discs 
is tuned below a critical value $t_c$,
 the system undergoes the superfluid-Mott insulator 
quantum transition.   
For a large number of atoms per well $(N \gg 1)$, 
the critical hopping rate can be evaluated within the hydrodynamic
approach \cite{wouters}. Neglecting the coupling with radial degrees of freedom and the particle loss due to inelastic processes,
the proper dynamical variables are the particle number fluctuation $N_\ell^\prime$ 
and the phase $\Phi_\ell$ of the order parameter in each disc. 
The hydrodynamic equations are equivalent to the classical equations of motion of the
1D phase Hamiltonian
\be\label{ham}
H_{P}=\sum_\ell (E_c/2){N_\ell^\prime}^2-E_J \cos(\Phi_{\ell+1}
-\Phi_\ell),
\ee
where $E_c=2\mu/N$ and $E_J=t^2 N/\mu$ are the charging and 
the Josephson energies, respectively, and  
$\hbar N_\ell^\prime$ and $\Phi_\ell$ are considered as
conjugated variables.

Quantization of the classical Hamiltonian (\ref{ham}) is 
achieved by replacing these variables with operators $\hbar \hat N^\prime$ and
$\hat \Phi$ satisfying the commutation relation 
$[\hbar \hat N^\prime,\hat \Phi]=i \hbar$. 
The quantized Hamiltonian is known to
exhibit a phase transition at the critical value 
$E_c =\eta E_j $, with $\eta \simeq 0.81$ \cite{bradley}.
The superfluid phase occurs for $E_c <\eta E_j $ and is characterized
by an algebraic decay of the phase correlation function 
$\langle \cos(\Phi_\ell-\Phi_k)\rangle$ at large distances 
$|\ell-k|\gg 1$.
The decay becomes exponential for $E_c >\eta E_j $ where one enters the Mott
regime, characterized by large phase fluctuations which
suppress interwell tunneling. 
The ground state is an insulator 
with a fixed number of atoms per disc and a finite gap in the excitation spectrum. 
By comparing the values of the charging and the Josephson 
energies, we find that for BCS superfluids 
\be\label{tc-bcs}
\frac{t_c}{\mu}= \frac{1}{N}\sqrt{\frac{2}{\eta}}.
\ee 
This result differs from the corresponding value for Bose 
condensates in the same geometry,  $t_c^{b}/\mu^{b}\sim 1/N^2$
where $\mu^{b}=ng_{2D}$ 
is the chemical potential and $g_{2D}$ is the 2D coupling constant.
This is because the Josephson energy in the 
Hamiltonian (\ref{ham}) for the bosonic case is $E_j^{b}=tN$.

Equation (\ref{tc-bcs}) has been derived under the assumption that
the effective tunneling rate $\nu_c=t_c^2/\hbar \mu$
is large compared to the loss rate $\tilde\nu$.
The most severe losses come from three-body recombination.
For an array of Bose-condensed atomic gases in the same geometry 
the corresponding loss rate is always large compared to 
the critical tunneling rate
\cite{dalibard}, unless the number of atoms per disc is  very small 
as in the experiment of Ref. \cite{greiner}. 
For Fermi superfluids the situation is completely different
because the inelastic processes are strongly inhibited by
quantum statistics. In the quasi2D geometry, in analogy with the 
3D case \cite{petrov1},
for the 3-body loss rate one can write
$
\tilde\nu=-\dot{n}/n=L n^2 (k_F R_e)^2,
$
where the small factor $(k_FR_e)^2$ comes from the Pauli principle,
$R_e$ is a characteristic radius of the interatomic potential, and $L$ 
is the quasi2D recombination coefficient. 
This coefficient is related to the corresponding quantity of a 3D
gas as $L\sim L_{3D}/\sigma^2$. Within an order of magnitude, $L_{3D}$
coincides with the recombination rate constant for bosonic isotopes of
the same atom, which ranges from $10^{-27}$ to $10^{-30}$cm$^6$/s for alkali
atoms. The ratio of the loss to the effective tunneling rate is then given by
\begin{equation}  \label{ratio}
\!\!\frac{\tilde\nu}{\nu_c}\approx\frac{\hbar L_{3D}n^2(k_FR_e)^2 N^2}{\mu\sigma^2}
\approx\left(\frac{2mL_{3D}}{\hbar R_e^2\sigma^2}\right)(nR_e^2)^2 N^2
\end{equation}
and should be much smaller than unity for consistency.

Note that in the case of bosons the effective tunneling rate is $\nu_c^b=t_c^b/\hbar\sim
ng_{2D}/N^2$. For a given $N$, it is smaller than the one for fermions by a factor of 
$(mg_{2D}/\pi\hbar^2)$, which is a small parameter of the theory for the 2D weakly interacting
gas. The bosonic 3-body loss rate is $\tilde\nu^b\sim (L_{3D}/\sigma^2)n^2$ and it
exceeds the fermionic loss rate by a factor of $(k_FR_e)^{-2}$. Thus, we have
$\tilde\nu/\nu_c\sim (mg_{2D}/\pi\hbar^2)(k_FR_e)^2\tilde\nu^b/\nu_c^b$.
For realistic densities the fermionic ratio $\tilde\nu/\nu_c$ is smaller than the bosonic ratio $\tilde\nu^b/\nu_c^b$
by 4 or 5 orders of magnitude. This is a consequence of Fermi statistics and
it is crucial for the observation of the Mott transition. 

For example, considering $N=10^3$ fermionic potassium atoms ($R_e\approx 5$ nm) in each
disc, with density $n=10^9$cm$^{-2}$ corresponding to a Fermi energy (chemical
potential) $\epsilon_F\simeq\mu=380$ nK k$_B$,  from 
Eq.(\ref{tc-bcs}) we find $t_c/\hbar \sim 70$ s$^{-1}$
corresponding to $s\sim 25$ for a lattice period $d=400$ nm, 
and $\sigma\approx 60$ nm. This leads to $\nu_c\sim 0.1$ s$^{-1}$ and, assuming
$L_{3D}\alt 10^{-28}$ cm$^6$/s, from Eq.(\ref{ratio}) we obtain
$\tilde\nu/\nu_c \alt 0.1$.
Hence, owing to quantum statistics, in Fermi superfluids one can easily 
have $\tilde\nu\ll\nu_c$ and achieve the Mott insulator transition. 
It is important to emphasize that, in the given example,
the suppression of recombination processes by a factor of 
$(k_FR_e)^2\sim 10^{-3}$ originating from the Pauli principle 
is crucial to keep the ratio (\ref{ratio}) small even for 
$N\sim 10^3$.

In conclusion, we have found the superfluid transition temperature for a two-component 
Fermi gas in a 1D optical lattice and revealed that the effective coupling constant 
depends in a non-trivial way on both the atom density and the parameters of the optical field.  
For an array of finite size discs with a large number of atoms per disc, we have shown that 
the critical effective tunneling rate for the Mott insulator quantum transition can be larger
than the rate of particles losses. Thus, 
the Mott phase can be observed for Fermi superfluids in this geometry. 

We acknowledge fruitful discussions with M.A. Baranov, L.P. Pitaevskii,  
M. Randeria, S. Stringari and W. Zwerger. We thank F. Dalfovo for a critical reading of the manuscript.
One of the authors (G. O.) wishes to thank the Van der Waals-Zeeman 
Institute (Amsterdam) and the LPTMS (Orsay) for the kind ospitality.
This work was supported by the Ministero dell'Istruzione, 
dell'Universita' e della Ricerca (M.I.U.R.), by the Minist\`ere de 
la Recherche (grant ACI Nanoscience 201), 
and by the 
Nederlandse Stichting voor Fundamenteel Onderzoek der Materie (FOM).

\end{document}